\documentclass[aps,twocolumn,pra,showpacs]{revtex4}
\usepackage{graphicx,amsmath,amssymb}

\newcommand{\ket}[1]{\left| #1\right\rangle}

\newcommand{\ketbra}[3][]{\left|#2\right\rangle_{#1}\!\left\langle#3\right|}
\newcommand{\ea}[0]{{\it et al. }}
\newcommand{\eac}[0]{{\it et al.}, }
\newcommand{\Tr}[0]{{\rm Tr}}
\newcommand{\id}{\mathbb{I}}
\newcommand{\eeqref}[1]{Eq.~(\ref{#1})}

\begin{document}

\title{Diluted maximum-likelihood algorithm for quantum tomography}

\author{Jaroslav \v{R}eh\'{a}\v{c}ek}
\email{rehacek@phoenix.inf.upol.cz}
\affiliation{Department of Optics, Palacky University, 17. listopadu 50,
77200 Olomouc, Czech Republic}
\author{Zden\v{e}k Hradil}
\email{hradil@optics.upol.cz}
\affiliation{Department of Optics, Palacky University, 17. listopadu 50,
77200 Olomouc, Czech Republic}
\author{E. Knill}
\email{knill@boulder.nist.gov}
\affiliation{Mathematical and Computational Sciences Division,
National Institute of Standards and Technology, Boulder CO 80305, USA}
\author{A. I. Lvovsky}
\email{lvov@ucalgary.ca}
\affiliation{Institute for Quantum Information Science, University of
  Calgary, Calgary, Alberta T2N 1N4,
  Canada}

\begin{abstract}
We propose a refined iterative likelihood-maximization algorithm for
reconstructing a quantum state from a set of tomographic measurements.
The algorithm is characterized by a very high convergence rate and
features a simple adaptive procedure that ensures likelihood increase in
every iteration and convergence to the maximum-likelihood state.
 We apply the algorithm to homodyne tomography of
optical states and quantum tomography of entangled spin states of
trapped ions and investigate its convergence properties.
\end{abstract}


\maketitle

\section{Introduction.}
Quantum tomography (QT) is a family of methods for reconstructing a
state of a quantum system from a variety of measurements performed on
many copies of the state. QT is of particular importance for quantum
information processing, where it is used to evaluate the fidelity of
quantum state preparation, capabilities of quantum information
processors, communication channels, and detectors. Theoretically
proposed in \cite{tomoprop} and first experimentally implemented in the
early 1990s \cite{smi93}, QT has become a standard tool in many branches
of quantum information technology.

Aside from the experimental procedure of conducting a set of
tomographically complete measurements on a system, QT requires a
numerical algorithm for extracting complete information about the state in
question from the measurement results. From a variety of algorithms
proposed, two main approaches have become popular among
experimentalists. One approach is based on linear inversion: because the
statistics of the measurement results is a linear function of the
density matrix, the latter can be obtained from the former by solving a
system of linear equations. Examples are the inverse Radon
transformation \cite{her80} or the quantum state sampling method
\cite{sampling}, that were almost exclusively used in optical homodyne
tomography until recently.

The second approach is \emph{maximum-likelihood} (MaxLik) quantum state
reconstruction, which aims  to find, among all possible
density matrices, the one which maximizes the probability of obtaining
the given experimental data set \cite{hradil97}.
To date, the maximum-likelihood approach has been applied to various
quantum problems from quantum phase estimation \cite{Qphase} to
reconstruction of entangled optical states \cite{Rehacek01,White01}.

MaxLik reconstruction has several advantages with respect to linear
inversion. First, with linear inversion, statistical and systematic
errors of the quantum measurements are transferred directly to the
density matrix, which may result in unphysical artifacts such as
negative diagonal elements. Second, MaxLik allows one to incorporate
additional information that may be known about the density matrix into
the reconstruction procedure. Third, experimental imperfections (such
as detector inefficiencies) can be directly incorporated in to the
MaxLik reconstruction procedure.


One approach to quantum MaxLik reconstruction is to
express the density matrix as a function of a set of independent
parameters, in a way that upholds the positivity and unity-trace
constraints for all parameter values. Then one can apply any iterative
optimization method to find the set of parameter values that maximize
the likelihood.  Because the log likelihood function for QT is convex,
the optimization problem is well behaved and most iterative
optimization methods are guaranteed to converge to the unique
solution.  This approach was used by James \ea in their work
on tomography of optical qubits \cite{White01}. In application to
homodyne tomography, the method was elaborated by Banaszek \ea
\cite{B5} and used in an experiment by D'Angelo \ea \cite{DAngelo}.

Generic numerical optimization methods are often slow when the number
of parameters (the square of the Hilbert space dimension) is large. An
alternative algorithm described below, which takes advantage of the
structure of the MaxLik reconstruction problem and has good
convergence properties was proposed by \cite{lnp} and later adapted to
different physical systems such as external degrees of freedom of a
photon \cite{vortex} and the optical harmonic oscillator \cite{mll}.
Thanks to its good properties, this method has been widely used in
recent experiments on optical homodyne tomography of both single- and
multimode optical states \cite{mllapp}. Despite its success, no
argument guaranteeing monotonic increase of the likelihood in every
iteration step has been presented.  Although to our knowledge the
experimental practice has not yet faced a counterexample,
theoretically such counterexamples do exist and there remains a risk
that the algorithm could fail for a particular experiment.

In this paper, we propose an iteration which depends on a single
parameter $\epsilon$ that determines the ``length'' of the step in the
parameter space. For $\epsilon\to\infty$, the iteration becomes that
of Ref. \cite{vortex,mll}. On the other hand, we prove that the
likelihood will increase in every iteration step for $\epsilon\to 0$.
We thus obtain a simple adaptive procedure, which, by choice of
parameter $\epsilon$, allows us to find a compromise between the
convergence rate and the guarantee on the likelihood increase.

\section{The nonlinear iterative algorithm.} \label{AlgSec} We now
describe the iterative scheme used in Refs. \cite{vortex,mll}.  A
generic tomographic measurement is described by a
positive-operator-valued measure
(POVM), with the outcome of the $j$'th measurement associated with a
specific positive operator $\hat\Pi_j\ge 0$, with $\sum_j \hat\Pi_j$ normalized to the identity operator.
In the case of sharp von Neumann measurements,
$\hat\Pi_j$ is a projection operator.

Let $N$ be the total number of
measured quantum systems and $f_j$ be the number of occurrences for each measurement result
 $\hat\Pi_j$. The
likelihood of a particular data set $\{f_j\}$ for the quantum state
$\hat{\rho}$ is given by $\mathcal{L}(\hat{\rho})=\prod_j{\rm pr}_j^{f_j}$,
with
\begin{equation}
{\rm pr}_j=\Tr[\hat\Pi_j\hat{\rho}]
\label{prj}
\end{equation}
being the probability of each outcome.

Our goal is to find the density matrix $\hat\rho$ which maximizes the
log-likelihood
\begin{equation} \label{Lgen}
\log \mathcal{L}(\hat{\rho}) = \sum_j f_j\log({\rm pr}_j).
\end{equation}
As was shown in Ref.~\cite{hradil97}, a state $\hat\rho_0$ that
maximizes the likelihood (\ref{Lgen}) obeys a simple nonlinear
extremal equation
\begin{equation}
\label{Rrho}\hat R(\hat{\rho}_0)\hat\rho_0=\hat\rho_0\hat
R(\hat{\rho}_0)=\hat\rho_0,
\end{equation}
where we introduced the state dependent operator
\begin{equation} \label{Rgen}
\hat R(\hat{\rho})=\frac{1}{N}\sum_j\frac{f_j}{{\rm pr}_j}\hat\Pi_j.
\end{equation}
Note that $R(\hat{\rho})$ is a non-negative operator.
Following Ref. \cite{Fiurasek01}, where a similar method was
proposed to estimate an unknown quantum measurement, Eq.~(\ref{Rrho})
can be stated in a slightly different but equivalent form
\begin{equation}  \label{RrhoR} \hat
R(\hat{\rho}_0)\hat\rho_0\hat R(\hat{\rho}_0)=\hat\rho_0.
\end{equation}
For simplicity, we assume that the measurements are sufficient
to ensure that there is a unique maximum likelihood state $\hat\rho_0$.

In the case where the density matrix $\hat\rho$ is restricted to
matrices that are diagonal, the problem of finding a solution to
Eq. (\ref{Rrho}) can be solved by the well-known
expectation-maximization algorithm~\cite{VardiLee}.  If $\hat R$ is
always diagonal in the same basis, expectation-maximization
reduces to computing the next iterate according to $\hat\rho^{(k+1)}=
\hat R(\hat\rho^{(k)}) \hat\rho^{(k)}$ in the hope of converging to a fixed
point that necessarily satisfies Eq. (\ref{Rrho}).  The
expectation-maximization algorithm is guaranteed to increase the
likelihood at every iteration step.  However, this iteration cannot be
used for the quantum problem because without the diagonal restriction,
it does not preserve the positivity of the density matrix. A possible
remedy is to apply the expectation-maximization iteration to the
diagonalized density matrix followed by a unitary transformation of
the density matrix eigenbasis~\cite{Rehacek01,Guta}.

Refs. \cite{vortex,mll} instead propose to base the iterative
algorithm on Eq.  (\ref{RrhoR}). We choose an initial density
matrix such as $\hat\rho^{(0)}=\mathcal{N}[\hat 1]$ (which avoids
any initial problems with zero ${\rm pr}_j$), and compute
the next iterate $\hat\rho^{(k+1)}$ from $\hat\rho^{(k)}$ using
\begin{equation}\label{iterhomo}
\hat\rho^{(k+1)}=\mathcal{N}\left[\hat
R(\hat\rho^{(k)})\hat\rho^{(k)}\hat R(\hat\rho^{(k)})\right],
\end{equation}
where $\mathcal{N}$ denotes normalization to trace $1$ and
the positivity of the density matrix is explicitly preserved in each step.
Hereafter we refer to scheme of
Eq.~(\ref{iterhomo}) as the ``$R\rho R$ algorithm''.

Despite the $R\rho R$ algorithm being a quantum generalization of the
well-behaving classical expectation-maximization algorithm, its
convergence is not guaranteed in general. This is evidenced by the
following counterexample. Assume that we made three measurements on a qubit with a
single apparatus with $\hat\Pi_0=\ketbra{0}{0}$,
$\hat\Pi_1=\ketbra{1}{1}$, detecting $\ket{0}$ once and $\ket{1}$
twice.  The measurement is tomographically incomplete because no
information is gained about the off-diagonal elements of the density
matrix.  From Eq.~(\ref{Rgen}), we find $\hat R=(\hat\Pi_0/\rho_{00} +
2 \hat\Pi_1/\rho_{11})/3$. Using the uniformly mixed
$\hat\rho^{(0)}=\hat\Pi_0/2+\hat\Pi_1/2$ as a starting point, we
obtain, in step 1:
\begin{equation}
\hat R =  \frac{2}{3} \hat\Pi_0 + \frac{4}{3} \hat\Pi_1;\qquad \hat
\rho^{(1)} =  \hat\Pi_0/5 + 4 \hat\Pi_1/5;
\end{equation}
and in step 2:
\begin{equation}
\hat R = \frac{5}{3} \hat\Pi_0 +
\frac{5}{6} \hat\Pi_1; \qquad
\hat \rho^{(2)} = \hat\rho^{(0)}.
\end{equation}
The iterations produce a cycle of length two. The second step strictly
decreases the likelihood.

\section{The ``diluted'' iterative algorithm.} To improve the
convergence of the $R\rho R$ iteration let us modify it along the lines
used for calculating the mutual entropy of entanglement in
\cite{mutual}, namely by mixing the generator of the nonlinear map
\eqref{RrhoR} with a unity operator
\begin{equation}\label{modified}
\hat\rho^{(k+1)}\equiv\hat I(\hat\rho^{(k)},\epsilon)=\mathcal{N}\left[\frac{\hat\id+\epsilon \hat
R}{1+\epsilon}\hat\rho^{(k)}\frac{\hat\id+\epsilon \hat R}{1+\epsilon}\right],
\end{equation}
where $\epsilon$ is a positive number. Loosely speaking, the nonlinear
map is diluted and the iteration step is controlled by $\epsilon$.
Now let us prove that using the modified algorithm \eqref{modified},
the likelihood is increased in each step if $\epsilon\ll 1$ is
sufficiently small.

In the linear approximation with respect to $\epsilon$, we can rewrite
Eq.~(\ref{modified}) as
\begin{equation}\label{modified2}
\hat\rho^{(k+1)}=\hat\rho^{(k)}+\Delta\hat\rho
\end{equation}
with
\begin{equation}
\Delta\hat\rho=\epsilon (\hat R\hat\rho^{(k)}+\hat\rho^{(k)}\hat
R-2\hat\rho^{(k)}).\label{modified2b}
\end{equation}
To obtain Eqs.~(\ref{modified2}) and (\ref{modified2b}), we approximated
$(1+\epsilon)^{-2}\approx 1-2\epsilon$ and used the relation
\begin{equation}\label{TrRrho}
\Tr(\hat R\hat\rho)=\Tr(\hat\rho\hat R)=1,
\end{equation}
which is a consequence of the definition (\ref{Rgen}) of $\hat R$.
The normalization factor $\mathcal{N}$ is $1$ to first order in $\epsilon$.

We now evaluate the likelihood associated with the new state
$\hat\rho^{(k+1)}$ and compare it to that of $\hat\rho^{(k)}$,
neglecting terms of second and higher order in $\epsilon$:
\begin{eqnarray}\label{longeq}
\log\mathcal{L}(\hat\rho^{(k+1)})&=&\sum f_j\log \Tr(\hat \Pi_j
\hat\rho^{(k+1)})\\ \nonumber
&=&\sum f_j\log [{\rm pr}_j+\Tr(\hat \Pi_j \Delta\hat\rho)]\\ \nonumber
&=&\sum f_j\log {\rm pr}_j+\log[1+\frac{1}{{\rm pr}_j}\Tr(\hat \Pi_j
\Delta\hat\rho)]\\ \nonumber
&=&\log\mathcal{L}(\hat\rho^{(k)})+\sum \frac{f_j}{{\rm pr}_j}\Tr(\hat \Pi_j
\Delta\hat\rho)\\ \nonumber
&=&\log\mathcal{L}(\hat\rho^{(k)})+\Tr(\hat R \Delta\hat\rho)\\ \nonumber
&=&\log\mathcal{L}(\hat\rho^{(k)})+2\epsilon [\Tr(\hat R \hat\rho^{(k)} \hat R)-1].\\
\nonumber
\end{eqnarray} In the second equality above, we used Eq.~(\ref{prj}); in
the fourth, the definition (\ref{Lgen}) of the likelihood and the
approximation $\log(1+\alpha)\approx\alpha$ for $\alpha\ll 1$; and in
the sixth, the cyclic property of the trace and Eqs.~(\ref{modified2b}) and
(\ref{TrRrho}).

We complete the proof by showing that
\begin{equation}
\mathrm{Tr}(\hat R\hat \rho \hat R)=\mathrm{Tr}(\hat R\hat\rho \hat
R)\mathrm{Tr}\hat \rho
\ge\mathrm{Tr}^2(\hat R\hat\rho)=1.\label{c-s-app}
\end{equation}
Indeed, the positive density matrix has a positive square root
$\hat\rho=(\hat\rho^{1/2})^2$, and thus $\mathrm{Tr}(\hat R\rho \hat
R)\mathrm{Tr}\hat \rho=(\hat R\hat\rho^{1/2} ,\hat R\hat
\rho^{1/2})(\hat\rho^{1/2}, \hat \rho^{1/2})$ and $\mathrm{Tr}^2(\hat
R\hat\rho)=|(\hat R\hat \rho^{1/2} ,\hat \rho^{1/2})|^2$, where the
scalar product of matrices is defined as $(\hat A,\hat B)
=\sum_{i,j}A^*_{ij}B_{ji}=\Tr(\hat A^\dag\hat B)$. Consequently the
Cauchy-Schwarz inequality can be applied to yield the inequality in
Eq.~(\ref{c-s-app}).

We have thus proven that under the application of iterations
\eqref{modified}, the likelihood is non-decreasing provided that
$\epsilon$ is chosen sufficiently small in every step.  Suppose we
find a density matrix $\hat \rho$ such that there is no
$\epsilon>0$ that yields a proper increase in the likelihood when
the iteration \eqref{modified} is applied. Then
\begin{equation}\label{extermecond}
\Tr(\hat R \hat\rho \hat R)=1.
\end{equation}
According to Eqs.~\eqref{TrRrho}, \eqref{c-s-app}, and the equality
condition in the Cauchy-Schwarz inequality, \eeqref{extermecond} can
be fulfilled if and only if $\hat R\hat\rho^{1/2}=\hat\rho^{1/2}$ or,
equivalently, $\hat R\hat \rho = \hat \rho$. The latter equality
characterizes the maximum likelihood state, so $\hat\rho=\hat\rho_0$.

Proper use of the diluted iterations requires a strategy for choosing
$\epsilon$ at each step. Asymptotic convergence of the diluted
iterations may depend on this strategy. As we show in the Appendix, one strategy that converges
to $\hat\rho_0$ is to choose the $\epsilon$ which maximizes the
likelihood increase in every iteration. However, this strategy is computationally expensive because it requires solving a one-dimensional optimization problem.

One possible alternative approach is as follows.
\begin{itemize}
\item  begin with
 the $R\rho R$ (\ref{iterhomo}) iterations, which are identical to the diluted
 iterations (\ref{modified}) with $\epsilon\to\infty$. Verify that the likelihood increases in each step.
\item In the event the likelihood does not increase and before
terminating the iterations, use the diluted iteration
(\ref{modified}), trying smaller values of $\epsilon$ to determine
whether significant increases in likelihood are still possible. If so,
continue the iterations as needed with these smaller values of
$\epsilon$.
\item When the iterations appear to have converged or stagnated, find the value of $\epsilon$ at which the likelihood increase is maximized and attempt additional iterations using this value. If the likelihood and/or the density matrix does not exhibit significant further changes, one can be sure the iteration sequence has converged to the maximum-likelihood solution.
\end{itemize}

Another approach is to choose $\epsilon$ randomly
according to a distribution with nonzero density in a neighborhood of
$0$. To ensure non-decreasing likelihood, each iteration requires
repeatedly choosing $\epsilon$ randomly until one is found for which
the likelihood increases. The argument in the Appendix can be expanded to show
that if $\epsilon$ is chosen in this way, then the iteration has a
non-zero probability of escaping from any non-maximum likelihood
density matrix.

We note again that in all practical cases studied so far the $R\rho R$
algorithm exhibited good convergence and monotonic increase of the
likelihood. The diluted iteration may become necessary for
low-dimensional systems where the nonlinear $R\rho R$ iteration may
``overshoot''. Characterizing the situations where this can happen
is an open problem.

Finally, let us mention that in some tomography schemes,
one or more POVM elements (measurement channels) $\Pi_j$
are not accessible and, consequently,
$\hat G\equiv \sum_j \hat\Pi_j$ may not be normalizable to the unity operator
on the reconstruction subspace.
Then the extremal map (\ref{RrhoR}) should be replaced by
$\hat G^{-1}\hat R(\hat \rho_0) \hat \rho_0
\hat R(\rho_0)\hat G^{-1}=\hat \rho_0$ to avoid
biased results, see e.g. \cite{Rehacek01}.
Obviously, the corresponding iterative procedure
can be diluted in a similar way as was done with the original
$R\rho R$ algorithm.

\section{Examples}
First, consider the counterexample discussed above. A simple
numerical test shows that replacing the $R\rho R$ iteration by
(\ref{modified}) warrants convergence for any finite $\epsilon$; the
likelihood monotonically increases for $\epsilon\lesssim 25.7$.

\begin{figure}[b]
  \center{\includegraphics[width=\columnwidth]{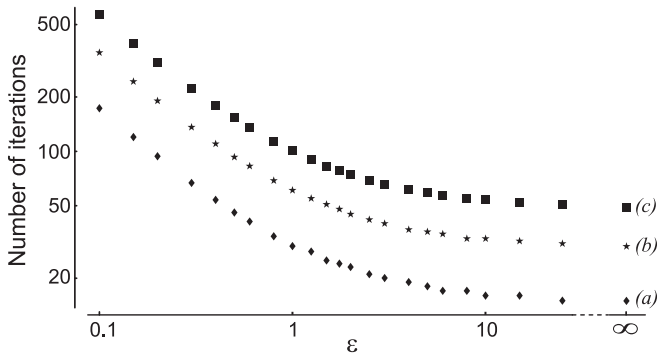}}
  \caption{Iterations required for  homodyne tomography data.
  The number of iterations required for convergence as a
  function of $\epsilon$. Results for three tolerances are shown:
  $10^{-3}$ (a),  $10^{-5}$ (b) and $10^{-7}$ (c). The rightmost column
  represents the $R\rho R$ algorithm ($\epsilon\to\infty$). The dataset
  is the same as that used in Ref.~\cite{mll}: 14,153 quadrature samples
  of the state approximating a coherent superposition of the vacuum and
  the single-photon states.}
  \label{fig:1}
\end{figure}

Second, we studied the dataset of 14,153 points obtained in the experiment
on homodyne tomography of the coherent superposition of
the vacuum and the single-photon Fock state \cite{catalysis}. This is
the same dataset as that analyzed in Ref.~\cite{mll}. This reference
discusses the specifics of application of the likelihood-maximization
procedure to continuous-variable measurements. We studied the
dependence of the convergence speed on the parameter $\epsilon$.

The Hilbert space was restricted to 14 photons. We first ran the
iterations for a very long time until the density matrix and the
likelihood no longer changed. In this way, we obtained the
density matrix $\hat\rho_0$ that maximizes the likelihood for this
dataset with high accuracy (limited by the floating point representation).

We then re-initialized the density matrix and ran the diluted $R\rho
R$ algorithm with various values of $\epsilon$.  We repeated the
iterations until the pairwise difference between all matrix elements of $\hat\rho^{(k)}$ and
$\hat\rho_0$ was below a pre-selected tolerance for each matrix
element. Three tolerance values were investigated: $10^{-3}$,
$10^{-5}$ and $10^{-7}$. The numerical experiment was conducted on a 2.8-GHz
Pentium 4 computer~\cite{tradenames}.
The code was written in Delphi~\cite{tradenames}. Each iteration
took about 0.3 s.

The result of this experiment is shown in Fig.~\ref{fig:1}.
The $R\rho R$ algorithm showed monotonic likelihood increase and
converged to the set tolerances within 15, 30, and 49 iterations,
respectively. The convergence rate of the diluted algorithm
improves with increasing $\epsilon$ and approached that of the $R\rho R$
algorithm for large values of $\epsilon$.  One sign of systematic
overshoot of the $R\rho R$ algorithm would be a minimum in the three
curves of Fig.~\ref{fig:1} at $\epsilon < \infty$. We did not observe
such an effect.

\begin{figure}[b]
  \center{\includegraphics[width=\columnwidth]{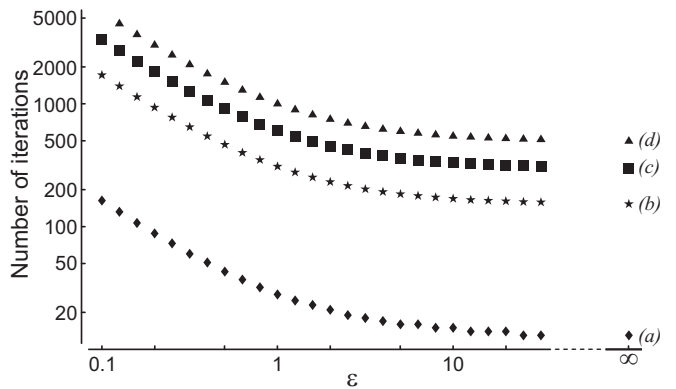}}
  \caption{Required iterations for ion-qubit tomography data.
   See the caption of Fig.~\ref{fig:1} for an explanation
   of the axes and interpretation of the curves. The tolerances used
   are $10^{-2}$ (a), $10^{-3}$ (b), $10^{-4}$ (c) and $10^{-7}$ (d).
   The underlying data was used in the analysis for~\cite{reichle:qc2006a}.}
  \label{fig:2}
\end{figure}

Third, we considered a dataset consisting of 21,832 individual
experiments with four ion qubits. The goal of the experiment was to
purify one entangled pair of ions from two~\cite{reichle:qc2006a}. In
order to determine the fidelity of the purified pair and for the
purpose of checking that the experiment did not introduce spurious
entanglement, incomplete state tomography was used. Specifically, each
tomographic measurement involved first determining the number of
qubits in state $\ket{0}$ among the first pair of qubits and then
performing a pair of $\pi/2$ pulses at various phases on the second
pair of qubits (the purified pair) before determining its number of
ions in state $\ket{0}$. Repetition of these combinations of pulses and
measurements suffices for determining the fidelity of the purified
entangled state.  The actual measurements involve counting the number
of photons scattered from an ion pair. This number has a Poissonian
distribution whose mean depends on the number of ions in state
$\ket{0}$. Thus, each experiment results in two counts, one from
each measurement. Every combination of counts can be associated with a
measurement operator of a POVM that also depends on the phases in the
pair of $\pi/2$ pulses. Although we cannot determine the density
matrix of the complete four-qubit state $\hat\rho$ with these
measurements, there is sufficient information to deduce the density
matrix $\hat\rho'$ obtained from $\hat\rho$ by phase decohering the
first pair of qubits in the logical basis and then symmetrizing each
pair of qubits.  The symmetrization process is equivalent to randomly
switching the qubits in each pair.  The diluted $R\rho R$ iteration
with the appropriate POVMs preserves the decohered and symmetrized
form of density matrices.  Starting from the completely mixed initial
state, it converges to the maximum likelihood solution for
$\hat\rho'$.  The code for the ion-qubit tomography was written in
R~\cite{Rref} and required about .3 s per iteration on a 1.6 GHz
Pentium 4 laptop.  The behavior of the iterations is shown in
Fig.~\ref{fig:2} and is similar to the behavior of the iterations for
the homodyne tomography shown in Fig.~\ref{fig:1}. Again, no sign of
overshoot was detected in these curves.

In summary, we have proposed an iterative likelihood-maximization
procedure for quantum tomography, which is applicable when the $R\rho
R$ iteration does not monotonically increase the likelihood. We have found the sufficient condition under which the iterations converge to the maximum-likelihood solution. The new
algorithm has been tested on two sets of experimental data.

\section*{Acknowledgements}
We thank J. Fiur\'{a}\v{s}ek for helpful discussions, D. Leibfried for
use of the ion trap data and S. Glancy and T. Gerrits for their help
in reviewing the paper.  This work was supported by NSERC, CFI, AIF, Quantum$Works$
and CIAR (A.L.); by the Czech Ministry of Education, Project
MSM6198959213, Czech Grant Agency, Grant 202/06/307  and the European Union project COVAQIAL FP6- 511004 (J.\v{R}
and Z. H.).  Contributions to this work by NIST, an agency of the US
government, are not subject to copyright laws.

\appendix

\section{Convergence of diluted iterations}

Consider the diluted iterations~(\ref{modified}).  For any fixed
$\epsilon$, we cannot exclude the possibility that the iteration
stagnates, even if the likelihoods continue to increase.  The problem
is that the direction of change in $\hat\rho$ for small $\epsilon$,
which is computed as a fixed function of $\hat\rho$, may differ
substantially from the direction of steepest ascent.  Although the
likelihood is guaranteed to increase for sufficiently small
$\epsilon$, the direction of change could become increasingly parallel
to surfaces of constant likelihood, thus leading to an iteration that
never reaches the maximum likelihood solution
$\hat\rho_0$. Alternatively, if $\epsilon$ is held fixed, the
iterations could converge into a limit cycle or a more complicated
limit set, thus avoiding $\hat\rho_0$.

Here we show that if at each step, $\epsilon$ is chosen to maximize
the likelihood increase, the iterations converge to $\hat\rho_0$ in
the limit.  To see this, we first notice that $\hat
R(\hat\rho)$ is continuous as a function of $\hat\rho$ on the set $S$ of density matrices for
which the likelihood is not $0$. This is because, if $\hat\rho\in S$, then ${\rm
pr}_j>0$ for all $j$ with $f_j>0$. It follows that the iterate $\hat I$
defined by Eq.~(\ref{modified}) is also a continuous function of the
density matrix and $\epsilon\geq 0$.

The initial density matrix $\hat\rho^{(0)}$ (the completely mixed state) is in $S$ because, for any $j$, ${\rm pr}_j(\hat\rho^{(0)})=\Tr[\hat\Pi_j\hat{\rho^{(0)}}]\propto\Tr[\hat\Pi_j]>0$. The choice of $\epsilon$ guarantees that the likelihood is non-decreasing, so each subsequent iterate $\hat\rho^{(k)}$ must be in $S$ as well. The sequence $\hat\rho^{(k)}$ is bounded and must thus have at least one limit point $\hat\rho_l$, which also belongs to the interior of $S$.

Suppose that
$\hat\rho_l\not=\hat\rho_0$. As we showed in the text, this implies
that the likelihood of $\hat I(\hat\rho_l,\epsilon)$ strictly
increases for sufficiently small $\epsilon$. In particular, there is
a $\delta>0$ and an $\epsilon$, such that the likelihood increase at
$\hat\rho_l$ is at least $\delta$.  Because the likelihood increase is
also a continuous function of $\hat\rho$ and $\epsilon$ on a
neighborhood of $\hat\rho_l$, there is a (possibly smaller)
neighborhood $S_\delta$ in which the maximum likelihood increase exceeds
$\delta/2$.  Because $\hat\rho_l$ is a limit point, one can choose an iterate
$\hat\rho^{(k)}$ in $S_\delta$ so that its
likelihood is within (say) $\delta/4$ of that of $\hat\rho_l$. Then
the next iterate's likelihood exceeds that of $\hat\rho_l$ by at least
$\delta/4$. Since the likelihood is non-decreasing, and by continuity,
future iterates cannot have $\hat\rho_l$ as a limit point,
contradicting the assumption on $\hat\rho_l$. We conclude that
$\hat\rho_l=\hat\rho_0$, as desired.



\begin{thebibliography}{99}
\bibitem{tomoprop}
R. G. Newton and  B. L. Young, Ann. Phys. (New York), {\bf 49}, 393
(1968); J. L. Park and W. Band, Found. Phys., {\bf 1}, 211 (1971); W.
Band, and J. L. Park, Am. J. Phys., {\bf 47}, 188 (1979); Found. Phys.,
{\bf 1}, 133 (1970); Found. Phys., 1, 339 (1971); J. Bertrand and P.
Bertrand, Found. Phys. {\bf 17}, 397 (1987); K. Vogel  and H. Risken,
Phys. Rev. A {\bf 40}, 2847 (1989).

\bibitem{smi93}
D. T. Smithey, M. Beck, M G. Raymer and A. Faridani, Phys. Rev. Lett. {\bf 70}, 1244 (1993);
D. T. Smithey,M. Beck, J. Cooper, M. G. Raymer and A. Faridani, Physica Scripta {\bf T48}, 35 (1993).

\bibitem{her80}  G. T. Herman, {\it Image Reconstruction from
Projections: The Fundamentals of Computerized Tomography} (Academic
Press, New York, 1980).

\bibitem{sampling}
G. M. D'Ariano, C. Macchiavello, and M. G. A. Paris, Phys. Rev. A {\bf
50}, 4298 (1994); G. M. D'Ariano, M. G. A. Paris, M. F. Sacchi, in
Quantum State Estimation, M. Paris and J. Rehacek (Eds.), Lect. Notes
Phys. 649 (Springer, Berlin Heidelberg, 2004); U. Leonhardt {\it et
al.}, Opt. Commun. {\bf 127}, 144 (1996).

\bibitem{hradil97}
Z. Hradil, Phys. Rev. A {\bf{55}}, R1561 (1997).

\bibitem{VardiLee} Y. Vardi and D. Lee, J. R. Statist. Soc B
{\bf 55}, 569 (1993).

\bibitem{Qphase}  J. \v{R}eh\'{a}\v{c}ek, Z. Hradil, M. Zawisky, S. Pascazio,
H. Rauch, and J. Pe\v{r}ina, Phys. Rev. A {\bf 60}, 473 (1999).

\bibitem{Rehacek01} J. \v{R}eh\'{a}\v{c}ek, Z. Hradil, and
M. Je\v{z}ek, Phys. Rev. A {\bf 63}, 040303(R) (2001).

\bibitem{White01} D.F.V. James, P.G. Kwiat, W.J. Munro, and A.G. White,
Phys. Rev. A {\bf 64}, 052312 (2001).

\bibitem{B5} K. Banaszek, G. M. D'Ariano, M. G. A. Paris, M. F. Sacchi,
Phys. Rev. A {\bf 61}, 010304(R) (1999).

\bibitem{DAngelo} M. D'Angelo, A. Zavatta, V. Parigi, M. Bellini,
quant-ph/0602150

\bibitem{lnp}
Z. Hradil, J. \v{R}eh\'{a}\v{c}ek, J. Fiur\'{a}\v{s}ek, and
M. Je\v{z}ek in \textit{Quantum State Estimation} edited by
M. Paris and J. \v{R}eh\'{a}\v{c}ek, Lect. Notes Phys. { \bf 649}
(Springer, Berlin Heidelberg, 2004).

\bibitem{vortex}
G. Molina-Terriza, A. Vaziri, J. Rehacek, Z. Hradil, A. Zeilinger,
Phys. Rev. Lett. {\bf 92}, 167903 (2004).

\bibitem{mll} A. I. Lvovsky, J. Opt. B: Q. Semiclass. Opt. {\bf 6}
S556 (2004).

\bibitem{mllapp} A.~Ourjoumtsev, R. Tualle-Brouri, P.~Grangier, Phys.
Rev. Lett. {\bf 96}, 213601 (2006); Science 312, 83 (2006); J. S.
Neergaard-Nielsen \eac quant-ph/0602198; S. A. Babichev, B. Brezger, A.
I. Lvovsky, Phys. Rev. Lett. {\bf 92}, 047903 (2004); S. A. Babichev, J.
Appel, A. I. Lvovsky, Phys. Rev. Lett. {\bf 92}, 193601 (2004).




\bibitem{Fiurasek01} J. Fiur\'{a}\v{s}ek, Phys. Rev. A {\bf 64},
024102 (2001).

\bibitem{Guta} L. M. Artilles, R. D. Gill, and M. I. Guta, J. R.
Statist. Soc B {\bf 67}, 109 (2005).

\bibitem{mutual}
J. Rehacek, Z. Hradil, Phys. Rev. Lett. {\bf 90}, 127904 (2003).

\bibitem{catalysis} A. I. Lvovsky and J. Mlynek, Phys. Rev. Lett.
{\bf 88} 250401 (2002).

\bibitem{reichle:qc2006a}  R. Reichle, D. Leibfried, E. Knill, J. Britton, R. B. Blakestad, J. D. Jost, C. Langer, R. Ozeri, S. Seidelin and D. J. Wineland,
Nature {\bf 443}, 838 (2006).


\bibitem{tradenames} The use of trade
names is for informational purposes only and does not imply endorsement
by NIST.


\bibitem{Rref} The R Project for Statistical Computing,
\texttt{http://www.r-project.org}.
\end{thebibliography}
\end{document}